# LOAD BALANCING WITH REDUCED UNNECESSARY HANDOFF IN ENERGY EFFICIENT MACRO/FEMTO-CELL BASED BWA NETWORKS

Prasun Chowdhury[1], Anindita Kundu[2], Iti Saha Misra[3], Salil K Sanyal[4]

Department of Electronics and Telecommunication Engineering, Jadavpur University, Kolkata-700032, India
[1]`prasun.jucal@gmail.com`, [2]`kundu.anindita@gmail.com`,
[3]`itisahamisra@yahoo.co.in`, [4]`s_sanyal@ieee.org`

## ABSTRACT

*The hierarchical macro/femto cell based BWA networks are observed to be quite promising for mobile operators as it improves their network coverage and capacity at the outskirt of the macro cell. However, this new technology introduces increased number of macro/femto handoff and wastage of electrical energy which in turn may affect the system performance. Users moving with high velocity or undergoing real-time transmission suffers degraded performance due to huge number of unnecessary macro/femto handoff. On the other hand, huge amount of electrical energy is wasted when a femto BS is active in the network but remains unutilized due to low network load. Our proposed energy efficient handoff decision algorithm eliminates the unnecessary handoff while balancing the load of the macro and femto cells at minimal energy consumption. The performance of the proposed algorithm is analyzed using Continuous Time Markov Chain (CTMC) Model. In addition, we have also contributed a method to determine the balanced threshold level of the received signal strength (RSS) from macro base station (BS). The balanced threshold level provides equal load distribution of the mobile users to the macro and femto BSs. The balanced threshold level is evaluated based on the distant location of the femto cells for small scaled networks. Numerical analysis shows that threshold level above the balanced threshold results in higher load distribution of the mobile users to the femto BSs.*

## 1. INTRODUCTION

The recent development of hierarchical macro/femto cell based broadband wireless access (BWA) networks is drawing the attention of wireless service providers more than ever due to their enhanced indoor coverage. The hierarchical network architecture not only provides extension in the cell coverage but also provides increase in the capacity and service quality enhancement. In BWA networks like WiMAX [1], femtocells are cost effective means to provide ubiquitous connectivity. The femto cellular base station is a miniaturized low-cost and low-power Base Station (BS) which uses a general broadband access network as its backhaul [2]. With the introduction of femto cells the total number of active users in the service area increases due to capacity enhancement. However, the mobility of these active users leads to huge number of macro/femto handoff in the hierarchical cell structures. On the other hand, if the load of the network is low most of the femto BSs remain unutilized even though it consumes power. Hence the power conservation of the entire hierarchical network along with elimination of unnecessary handoff provides a significant area of research work.

In recent literatures like [3], [4] authors have presented WiMAX femto cell system architectures and evaluate its performance in terms of network coverage, system capacity and performance of mobile station in indoor environment. On the other hand, authors of [5], [6] have compared the performance in private and public access method in WiMAX femto cell environment. However, the process of handoff, QoS requirement of the mobile stations and reducing the energy consumption of the femto BSs have not been considered in any of the aformentioned papers.

A variety of handoff algorithms based on received signal strength (RSS) have been considered in [7], [8], [9]. Velocity of the mobile node have been considered in [8], [9] as a parameter for handoff decision. However, QoS guarantee for the real time service and conservation of energy of the hierarchical network has not been considered in these papers too.

In [10] authors have proposed a new handoff algorithm to correctly assign the users to the femto cells but QoS profile, energy conservation and network load balancing are not taken into account. Moreover, no relation between the femto cells and the whole system has been considered in their simulation which does not reflect a real mobile WiMAX architecture.

In this paper, we have considered a WiMAX system where a mobile station is moving from a macro cell to a femto cell. Since at this stage the distance of the mobile station from the macro BS is more than the femto BS, more power will be consumed by the mobile station to communicate with the macro BS than the femto BS. Hence, we have assumed that the mobile station gives higher priority to the femto BSs than the macro BS. Thus a mobile staion selects the femto BS as its serving BS when it receives siganl from both the macro and femto BSs as well as the RSS from macro BS falls below its threshold level. In this paper, we have also determined the balanced threshold level of RSS from macro BS based on the distant location of the femto cells for small scaled network. Balanced threshold level provides equal load distribution of the mobile users to the macro and femto BSs. Numerical analysis shows that threshold level above the balanced threshold results in higher load distribution of the mobile users to the femto BSs.

To achieve QoS aware hierarchical networks, our handoff decision algorithm is based on two main factors viz. the velocity of mobile station and the service type of the mobile station. When a user moves with an ongoing call at a very high velocity (above velocity threshold) from one end of the hierarchical cell to the other end, it is expected that the user will experience huge number of macro/femto handoff in a short period of time. This burdens the overhead of the macro BS. A considerable amount of packet loss may also be encountered which degrades the call quality. On the other hand, though a user moves with a velocity lower than the velocity threshold it experiences comparitively lower handoff rate but handoff still happens. In this case, if the ongoing call is a real-time service then packet loss will hamper the call quality.

To avoid this quality degradation due to multiple number of macro/femto handoff, in our handoff decision algorithm we have considered a velocity threshold such that any user moving with a velocity higher than the velocity threshold or undergoing real-time transmission will not undergo any macro/femto handoff. Thereby, we eliminate unnecessary handoff and provide reduced network overhead and increased user satisfaction.

However, with the introduction of the femto cells the power consumption of the entitre network increases. The femto BSs may consume power even if no end user resides under its coverage. This may lead to wastge of power. To overcum this wastage of power we have considered a power conservation scheme by which the femto BSs will remain in a low power 'Idle mode' [11] with the pilot transmissions and processing switched off when no user is present in its coverage. Based on this considerartion our proposed handoff decision algorithm has been analyzed using Continuous Time Markov Chain (CTMC) Model.

The remainder of the paper is organized as follows: Section 2 discusses system model and the detailed description of the proposed energy efficient handoff decision algorithm of WiMAX macro/femto-cell networks. Section 3 shows analytical model and QoS performance evaluation parameters. Numerical results are discussed in section 4. Finally, Section 5 concludes the paper.

## 2. SYSTEM MODEL AND PROPOSED ENERGY EFFICIENT HANDOFF DECISION ALGORITHM

A WiMAX macro cell of 1.2 km radius is considered along with multiple femto cells deployed randomly at a distance of atleast 'R' meter from the macro BS as shown in Figure 1.

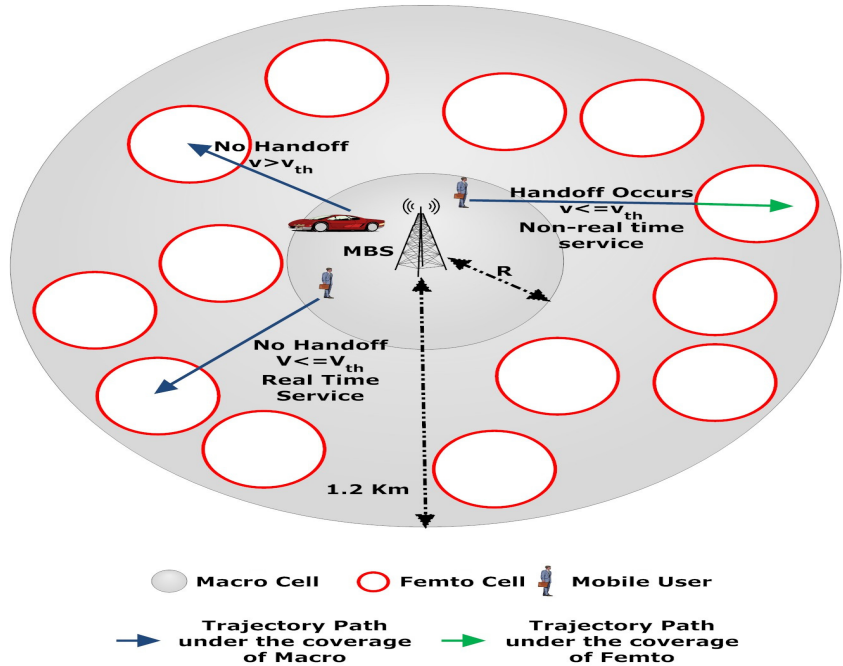

Figure 1. Energy efficient handoff decisions in the hierarchical system model

No femto cell is considered within 'R' meter of radius of the macro BS because the RSS of the mobile nodes residing within this area is assumed to be quite high. A number of mobile users are deployed randomly under the coverage of the macro BS with varied velocity and undergoing calls of varied service type. The rest of the system model parameters are shown in Table 1.

TABLE 1.  SYSTEM MODEL PARAMETERS

| Parameters | Value |
|---|---|
| Macro cell radius | 1.2 Km |
| Femto cell radius | 30m |
| Real-time service type | UGS, rtPS |
| Non real-time service type | nrtPS, BE |

Initially, the femto BSs are considered to be in the idle mode with all the pilot transmissions and associated radio processing disabled. The femto BSs incudes a low power sniffer ($P_{sniff}$) [11] which allows the detection of an active call originating from a mobile under its coverage to the macro BS. At this stage, the femto BS changes to active mode and requests the macro BS to handoff the newly originated call to it. Thus, the femto BSs are active only when any end user is active under its coverage which thereby enhancing the energy conservation.

Macro/femto cellular handoff comprises of two main phases – handoff strategy and handoff decision algorithm. The first phase deals with the RSS comparison while in the second phase the system decides when to trigger the handoff. In this paper, we contribute an efficent algorithm for the second phase. The handoff strategy proposed in [7] has been considered for the first phase.

Let $RSS_m$ and $RSS_f$ denote the received signal strength from the macro BS and femto BS experienced by a mobile node at any instant of time. As a mobile node moves in a straight line from the macro BS to femto BS with constant low velocity as shown in Figure 1, conventional handoff occurs. The conventional handoff algorithm with the RSS comaparison [7] can be expressed as in equation (1).

$$RSS_m < RSS_{m,th} \quad \text{and} \quad RSS_f > RSS_m + \Delta \tag{1}$$

where $RSS_{m,th}$ and $\Delta$ denotes the minimum RSS threshold level from the macro BS and the value of hysteresis respectively.

The pathloss encountered by a mobile node as it moves away from the macro BS diminishes the $RSS_m$. As the distance from the macro BS increases, this pathloss triggers the handoff situation where $RSS_f$ becomes higher than $RSS_m$. In our scenario, we have considered the ITU pathloss model in slow fading channel [12] as shown in equations (2) and (3).

$$PL_m = 15.3 + 37.6\log_{10}(D) + PL_{hw} \quad \text{where,} \quad PL_{hw} = 10 \tag{2}$$

$$PL_f = 38.46 + 20\log_{10}(d) + 0.7d \tag{3}$$

where $PL_m$ and $PL_f$ denotes the pathloss from macro and femto BS respectively. 'D' and 'd' are the corresponding distance of the mobile user from macro and femto BS.

Thus the resulting $RSS_m$ and $RSS_f$ encountered by the mobile node are shown in equations (4) and (5) respectively.

$$RSS_m = P_{m,tx} - PL_m \tag{4}$$

$$RSS_f = P_{f,tx} - PL_f \tag{5}$$

where $P_{m,tx}$ and $P_{f,tx}$ denote the transmit power of macro BS and femto BS respectively.

However, in our proposed handoff decision phase the conventional handoff is not adopted in the following two cases.

Case I: When the mobile user is moving at a very high velocity

Conventional handoff is not applicable when a mobile user is moving with a very high velocity. As a user moves with a very high velocity it undergoes huge number of macro/femto handoff within a very short period of time. The overhead of the macro BS thus increases unnecessarily.

Hence, in this paper we have considered a velocity threshold '$V_{th}$' of 10 Kmph (non motor vehicular speed) and simulated our scenario accordingly. If a user moves with a velocity 'V' such that V> $V_{th}$, unlike conventional scenario the user will not undergo handoff. Thus the unnecessary handoff is eliminated and improved QoS is guaranteed.

Case II: When the mobile user is undergoing a UGS or rtPS i.e. real-time call

When a user is moving with a real-time connection the number of handoff encountered degrades the call quality proportionately. Hence, in our scenario, no handoff is triggered for them in order to maintain the call quality. Thereby unnesecessary handoff count decreases and improved QoS is assured to the real-time users.

In our scenario, the QoS guarantee achieved by considering only case I is refered to as soft QoS guarantee while QoS guarantee achieved by considering both case I and case II is called hard QoS guarantee. Soft QoS guarantee only will reduce the overhead of the network. On the other hand, hard QoS guarantee will reduce the network overhead as well as increase user satisfaction.

## 3. ANALYTICAL MODEL AND PERFORMANCE EVALUATION PARAMETERS

In this paper, the performance evaluation of the WiMAX macro/femto-cell networks is obtained by using Continuous Time Markov Chain (CTMC) Model [13]. In addition, we have considered Pareto distribution for the arrival process of the priority service types, so the network model experiences a flow of service requests in a continuous time domain. The network undergoes a continuous change in its current state due to the occurrence of events (i.e. arrival and service of priority calls). It is necessary to observe the short-lived states of the network in order to analyze its performance more accurately. This is only possible if the network is modelled with CTMC.

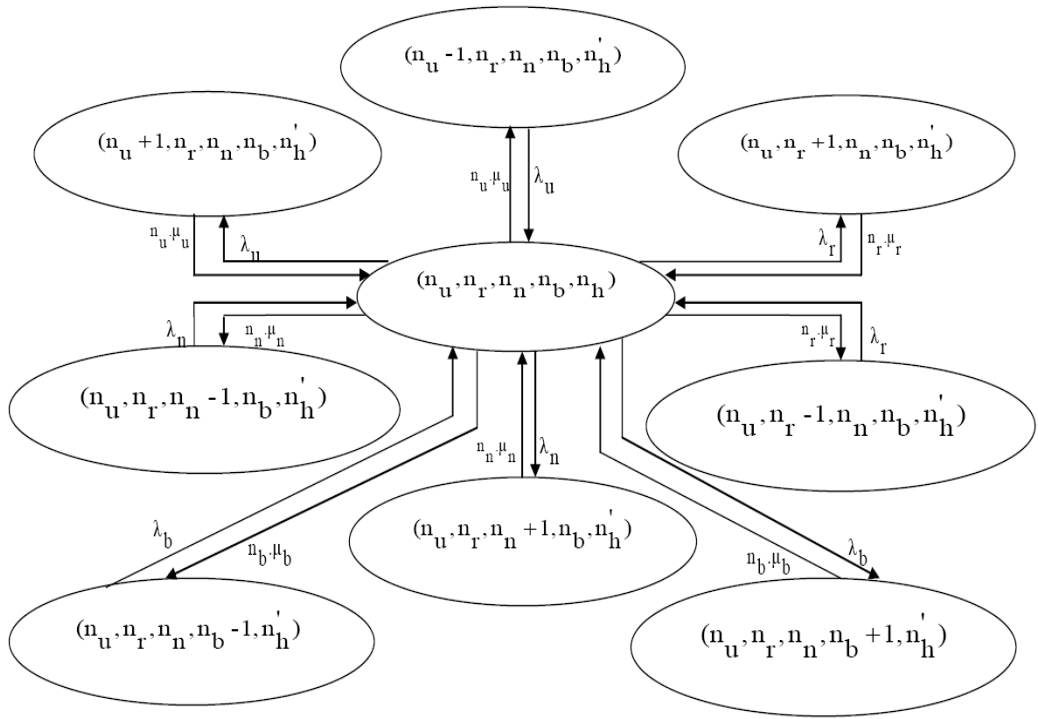

Figure 2. State transition diagram of the hierarchical WiMAX networks

International Journal of Wireless & Mobile Networks (IJWMN) Vol. 4, No. 3, June 2012

A hierarchical WiMAX networks consisting of single macro BS along with multiple femto BS is considered. The macro BS will receive the handoff requests from the users directly. Four types of services i.e. UGS, rtPS, nrtPS and BE need QoS guarantees and request for a handoff whenever it finds any suitable femto BS in the near vicinity. The hierarchical networks change state from one to another upon the admission or termination of a service type. Further, it is assumed that the hierarchical networks either admit or terminate only one service type at a particular instance of time. So the next state of the hierarchical networks depends only on the present state of the hierarchical networks but does not depend on the previous states of the hierarchical networks. Therefore, the states of the hierarchical networks form a Markov Chain and accordingly the hierarchical networks can be analytically modelled as shown in Figure 2. In this scenario the hierarchical networks can uniquely be represented in the form of a five dimensional Markov Chain $(n_u, n_r, n_n, n_b, n_h)$ based on the number of services residing within the hierarchical networks and the total number of macro/femto handoff occurred in the network.

State $s = (n_u, n_r, n_n, n_b, n_h)$ represents that the hierarchical networks have currently admitted '$n_u$', '$n_r$', '$n_n$' and '$n_b$' number of UGS, rtPS, nrtPS and BE service respectively. '$n_h$' represents the total number of macro/femto handoff occurred in that state of the hierarchical networks. We have assumed that initially no users are present under the coverage of femto cells. Hence, the total number of users present under the coverage area of femto BSs is also indicated by the parameter '$n_h$'. In Figure 2, $n_h'$ is the modified values of the variable '$n_h$' after state transition. Pareto distribution is considered for the arrival process of the newly originated UGS, rtPS, nrtPS and BE with rates of $\lambda_u, \lambda_r, \lambda_n$ and $\lambda_b$ respectively. This is because Pareto distribution supports more practical traffic model [14]. However, Poisson distribution is an ideal model, which is not practical in real WiMAX networks. The service times of UGS, rtPS, nrtPS and BE connections are exponentially distributed with mean $1/\mu_u, 1/\mu_r, 1/\mu_n$ and $1/\mu_b$ respectively.

Let the steady state probability of the state $s = (n_u, n_r, n_n, n_b, n_h)$ be represented by $\pi_{(n_u, n_r, n_n, n_b, n_h)}(s)$. As the Markov chain is irreducible, thereby observing the outgoing and incoming states for a given state '$s$', the steady state probabilities of all states of the hierarchical networks have been evaluated.

From the steady state probabilities we can determine various QoS performance parameters of the system as given below.

### 3.1. Handoff Probability (HO_Prob)

Number of handoff occurred in a particular state of the hierarchical networks multiplied with the steady state probability of that state will give the handoff probability of that particular state. Thereby, handoff probability of the hierarchical networks is obtained by summing the handoff probabilities of all the states of the hierarchical networks.

Hence, the probability of handoff occurred in hierarchical macro/femto networks can be calculated as follows.

$$\text{HO\_Prob} = \sum_{\forall s} n_h * \pi_{(n_u, n_r, n_n, n_b, n_h)}(s) \tag{6}$$

### 3.2. Macro Load (ML)

The Macro load is defined as the ratio of the number of users residing under macro BS to the total number of users present within the macro/femto hierarchical networks. ML can be calculated as follows.

$$ML = \sum_{\forall s} \frac{(n_u + n_r + n_n + n_b - n_h) * \pi_{(n_u, n_r, n_n, n_b, n_h)}(s)}{n_u + n_r + n_n + n_b} \quad (7)$$

### 3.3. Femto Load (FL)

The Femto load is defined as the ratio of the number of handoff users residing under femto BSs to the total number of users present within the macro/femto hierarchical networks. FL can be calculated as follows.

$$FL = \sum_{\forall s} \frac{n_h * \pi_{(n_u, n_r, n_n, n_b, n_h)}(s)}{n_u + n_r + n_n + n_b} \quad (8)$$

### 3.4. Energy consumption with active-idle mode ($E_{active\text{-}idle}$)

The probability that a femto BS is in active state is directly proportional to the network load and macro to femto handoff probability. In our markov chain model as we have considered exponential distribution of the service time of the users, hence the probability that a femto BS is in active state can be calculated as follows

$$\Pr ob(active) = (1 - e^{-\rho}) * HO\_prob \quad (9)$$

Where $\rho = \lambda/\mu$ for a particular service type and is termed as network load [15].

With the above consideration, probability that a given femto BS is in idle state is given as follows

$$\Pr ob(idle) = 1 - \Pr ob(active) \quad (10)$$

The power consumption in active state and idle state are denoted as '$P_{active}$' and '$P_{idle}$' respectively. The power of '$P_{sniff}$' is also additionally consumed in idle state. Hence, monthly energy consumption of a single femto BS in kWh with active-idle mode is given as follows

$$E_{active-idle} = (P_{active} * \Pr ob(active) + (P_{idle} + P_{sniff}) * prob(idle)) * 3600 * 24 * 30 \quad (11)$$

### 3.5. Energy consumption with conventional mode ($E_{conventional}$)

Since femto BSs are always remain in active state in the conventional mode so in this case the monthly energy consumption in kWh is given as follows

$$E_{conventional} = P_{active} * 3600 * 24 * 30 \quad (12)$$

## 4. NUMERICAL RESULTS AND DISCUSSIONS

The contribution of this paper lies in balancing the network load between the macro BS and femto BSs and the reduction of the unnecessary handoff at minimal energy consumption. Exhaustive simulations have been carried out under MATLAB version 7.3. Since our main aim is to reduce unnecessary handoff not the handoff latency, in our simulation the value of hysteresis has been taken as zero. The arrival rates of all the connections are assumed to be same i.e. $\lambda_u = \lambda_r = \lambda_n = \lambda_b$. The values of the rest of the simulation parameters are shown in Table 2. In Table 2 the macro and femto transmit power has been taken from [16]. In ideal case scenario, it is assumed that there is no instrumental power loss in the femto cells and hence we have considered that transmit power is equal to their input active power i.e. 20dBm=100mW. The results associated with the load balancing are shown in Figure 3, 4 and 5 while reduction of the unnecessary handoff is exhibited in Figure 6. Also, the monthly energy consumption for a single femto BS with the proposed handoff reduction technique is given in Figure 7, 8 and 9. Justifications behind all the numerical results have also been provided.

TABLE 2. SIMULATION PARAMETERS

| Parameters | Value |
|---|---|
| Macro transmit power ($P_{m,tx}$) | 46dBm |
| Femto transmit power ($P_{f,tx}$) | 20dBm |
| Femto active power ($P_{active}$) | 100mW |
| Femto idle power ($P_{idle}$) | 60mW |
| Low power sniffer ($P_{sniff}$) | 3mW |
| Velocity threshold ($V_{th}$) | 10kmph |
| $\mu_u = \mu_r = \mu_n = \mu_b$ | 0.2 |
| Traffic ratio of UGS, rtPS, nrtPS, BE | 1:1:1:1 |
| Femto cell deployment | Random |

As the threshold level of $RSS_m$ or $RSS_{m,th}$ increases, the load of the macro BS decreases and the load of the femto BSs increases. This is revealed from Figure 3a and 3b respectively. With increase in the value of $RSS_{m,th}$ the macro/femto handoff count increases. Hence, the macro load decreases while increasing the femto BSs load.

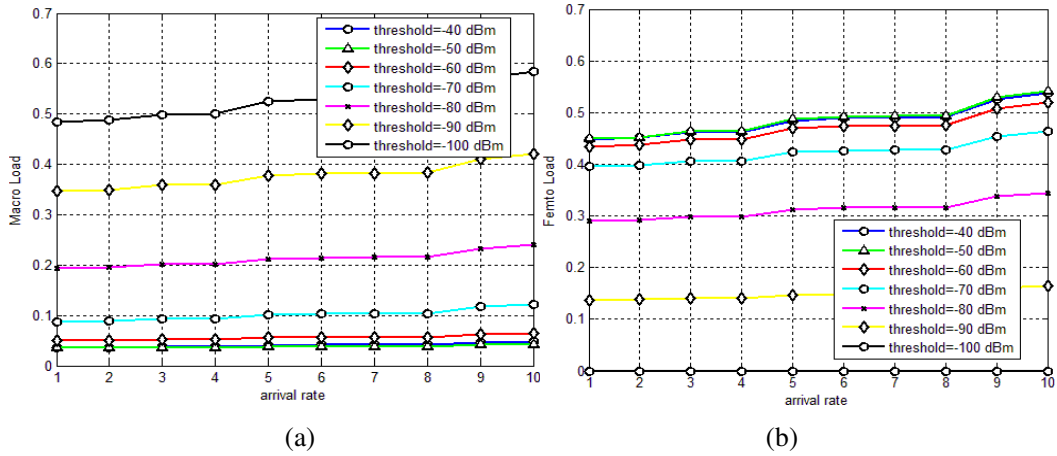

(a)  (b)

Figure 3. (a) Macro Load for various $RSS_{m,th}$ when 'R'=100m and (b) Femto Load for various $RSS_{m,th}$ when 'R'=100m

A saturation is observed in the load of both macro and femto BSs when $RSS_{m,th}$ reaches -50 dBm. No change is observed when value of $RSS_{m,th}$ increased further. It happens due to absence of femto cells within 'R'=100 m radius of macro BS as shown in Figure 1. Hence, when $RSS_{m,th}$ goes above -50 dBm, the mobile nodes residing within 'R'=100 m of macro BS do not find any femto BS. So no handoff is triggered and the load remains unchanged. Again, femto load is found to be zero at $RSS_{m,th}$= -100dBm due to macro cell outage. At this level of RSS theshold no femto cells are present to trigger handoff.

To balance the load of the macro BS and femto BSs, a study of the variation of their load has been performed with respect to the $RSS_{m,th}$. This variation is performed when femto cells are located at 'R'=100m apart from macro BS and is shown in Figure 4. The point of intersection of these two variations provides the value of $RSS_{m,th}$ at which the macro and femto load are observed to be same. It is observed that at $RSS_{m,th}$ =-83.4 dBm a balance between the load of the macro BS and femto BSs is achieved. Henceforth, Macro threshold level ($RSS_{m,th}$) at which load balancing is achieved is referred to as balanced threshold level. Load distribution to the femto BSs increases as the $RSS_{m,th}$ goes above the balanced threshold level i.e. -83.4 dBm.

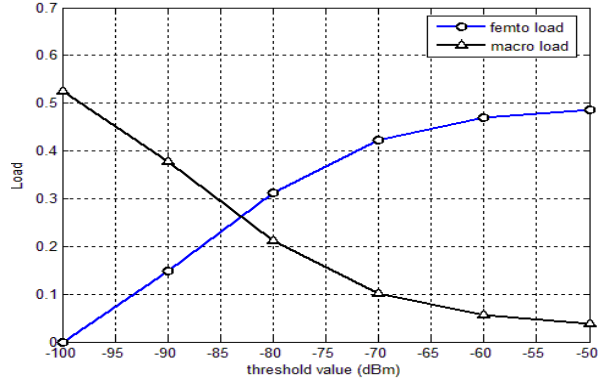

Figure 4. Comparison of macro/femto load for various $RSS_{m,th}$ when 'R'=100m

The above mentioned scenario has been generalized by varying the parameter 'R'. The corresponding balanced threshold level is evaluated in the similar way and is shown in Table 3. Thus, in order to have higher load distribution to the femto cells the macro BS should set the macro threshold level above the balanced threshold level with respect to the value of 'R'. The method of determining the balanced threshold level has been assessed for small scaled networks as the simulations are very time-consuming for broad scaled networks. This method can also be applied for broad scaled networks conceptually to calculate the corresponding balanced threshold level.

TABLE 3. THRESHOLD LEVEL FOR LOAD BALANCING

| Variation of 'R' (meter) | Balanced threshold level (dBm) |
|---|---|
| 200 | -85.7 |
| 300 | -87.1 |
| 400 | -88.6 |
| 500 | -89.8 |
| 600 | -90.9 |
| 700 | -92.1 |
| 800 | -93.8 |
| 900 | -95 |
| 1000 | -95 |
| 1100 | -95 |

In Table 3, as the value of 'R' increases, the balanced threshold level is observed to decrease gradually unless 'R' reaches 900 meters. From 900 meters onwards the macro threshold level is observed to remain constant at -95 dBm. The reason behind this is elaborated in Figure 5. Figure 5 shows the variation of the RSS of the mobile station with respect to its distance 'D' from the macro BS as obtained from equation (4). From Figure 5 we see that as the mobile station reaches the outskirt of the macro cell edge i.e. when D= 1.2 km, the RSS encountered is -95 dBm. Thus combining Figure 5 and Table 3 we conclude that even if the distance of the femto cells from the macro BS is 900 meters and beyond, the macro threshold has to be above -95 dBm to achieve higher load distribution to the femto cells.

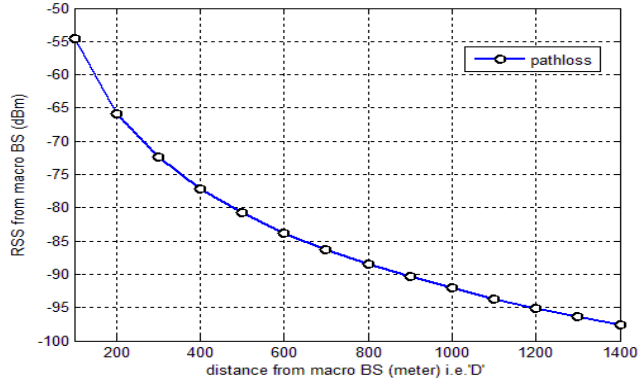

Figure 5. Pathloss from macro BS

Considering the above fact, we have observed the handoff probability for various kind of handoff decision discussed in this paper keeping $RSS_{m,th}$ at -70 dBm which is above the balanced threshold level for any 'R'. The result is shown in Figure 6.

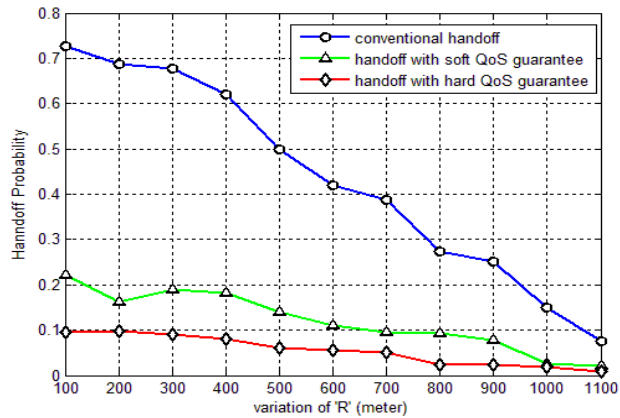

Figure 6. Handoff probability at $RSS_{m,th}$=-70dBm

Figure 6 reveals that the probability of handoff decreases to a considerable amount when unnecessary handoff is eliminated from the conventional handoff scheme. Again, the handoff probability in each case is also observed to fall gradually as 'R' increases. With increase in 'R', the outskirt region to be covered by the femto cells decreases. This in turn lowers the number of femto cells and thereby the handoff probability decreases. The handoff probability is observed to be much lower in case of hard QoS guarantee than the case of soft QoS guarantee. In a hierarchical cell scenario as shown in Figure 1, if a user moves with very high velocity or undergoes real-time call while moving from one end of the cell to the other end, reduction in the number of handoff is going to improve the call quality considerably. Thus soft QoS guarantee will ensure better call quality while hard QoS guarantee will provide much better call quality compared to conventional handoff scheme.

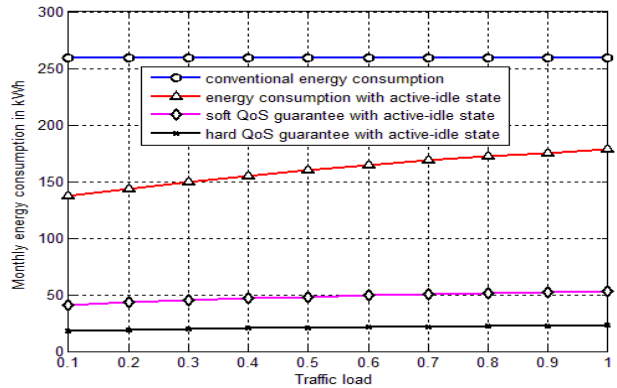

Figure 7. Monthly energy consumption for a single femto BS

Figure 7 shows the monthly energy consumption of the proposed handoff reduction technique considering the active-idle state of femto BSs with respect to the traffic load. As observed from the figure, the energy consumption remains unaffected with the variation of the traffic load for the conventional handoff strategy without consideration of the concept of active-idle state. However, the introduction of the active-idle state in the femto BSs conserves the monthly energy consumption to a huge extent. Energy is further conserved for the soft QoS and hard QoS guaranteed handoff reduction technique with the consideration of the active-idle state of femto BSs. Thus, the huge amount of energy is conserved by introduction of the active-idle mode of the femto BSs. The amount of energy saved is illustrated in Table 4.

TABLE 4. PERCENTAGE OF ENERGY SAVING

|  | Percentage of energy saving w.r.t conventional energy consumption | | | | |
|---|---|---|---|---|---|
| Traffic Load | 0.2 | 0.4 | 0.6 | 0.8 | 1.0 |
| With active-idle state | 47.81 | 43.70 | 40.33 | 37.58 | 35.32 |
| Active-idle state with soft QOS | 85.06 | 83.88 | 82.92 | 82.13 | 81.48 |
| Active-idle state with hard QOS | 92.76 | 92.19 | 91.72 | 91.34 | 91.03 |

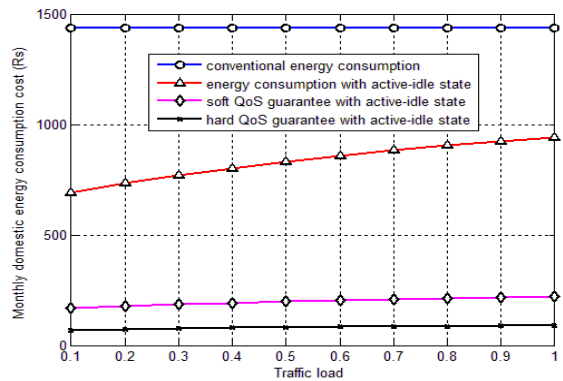

Figure 8. Monthly domestic energy consumption cost for a single femto BS

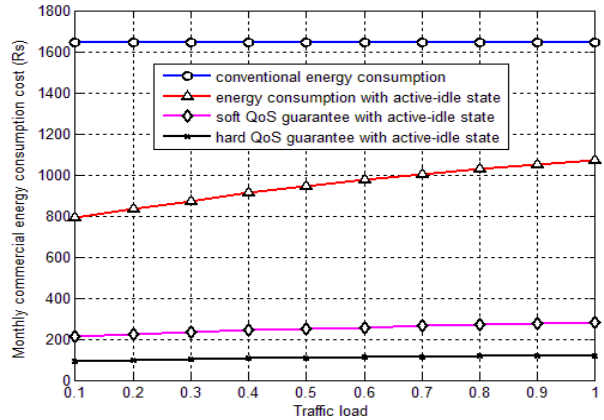

Figure 9. Monthly commercial energy consumption cost for a single femto BS

Based on the energy consumption tariffs of Calcutta Electricity Supply Corporation (CESC) of 2012 [17] the monthly energy consumption cost is calculated for domestic and commercial deployment of femto BS. Figure 8 and 9 reflects the domestic and commercial cost respectively for various handoff reduction techniques with active-idle state of femto BS. The profit gained for both the cases are summarized in Table 5 and 6.

TABLE 5.  DOMESTIC PROFIT GAINED

|  | Percentage of domestic profit gained w.r.t conventional energy consumption | | | | |
|---|---|---|---|---|---|
| Traffic Load | 0.2 | 0.4 | 0.6 | 0.8 | 1.0 |
| With active-idle state | 52.62 | 48.10 | 44.39 | 41.36 | 38.88 |
| Active-idle state with soft QOS | 88.99 | 88.06 | 87.30 | 86.67 | 86.16 |
| Active-idle state with hard QOS | 94.88 | 94.47 | 94.14 | 93.87 | 93.65 |

TABLE 6.  COMMERCIAL PROFIT GAINED

|  | Percentage of commercial profit gained w.r.t conventional energy consumption | | | | |
|---|---|---|---|---|---|
| Traffic Load | 0.2 | 0.4 | 0.6 | 0.8 | 1.0 |
| With active-idle state | 52.65 | 48.43 | 44.83 | 41.77 | 39.26 |
| Active-idle state with soft QOS | 87.70 | 86.73 | 85.94 | 85.29 | 84.76 |
| Active-idle state with hard QOS | 94.04 | 93.57 | 93.19 | 92.87 | 92.61 |

## 5. CONCLUSIONS

In this paper, we have proposed an energy efficient handoff decision algorithm for reducing unnecessary handoff in hierarchical macro/femto networks while balancing the load of macro and the femto BSs at minimal energy consumption. The performance of the proposed algorithm is also analyzed using CTMC model. Balanced threshold level of RSS from macro BS have been evaluated with respect to the distant location of the femto cells. Macro threshold level ($RSS_{m,\,th}$) set above the balanced threshold level results in higher load distribution of the mobile users to the femto cells. Hard QoS and soft QoS guarantee – the two decision of handoff reduction technique proposed in this paper shows how unnecessary handoff has been reduced while balancing the load of the macro and femto BSs. Soft QoS guarantee only will reduce the overhead of the network. On the other hand, hard QoS guarantee will reduce the network overhead as well as increase user satisfaction. So a service provider can choose any one of the QoS guarantee level depending upon their requirement. In addition, introduction of the active-idle state of the femto BSs shows significant reduction in the monthly energy consumption which is reflected in the domestic and commercial energy cost.

Since simulations are very time-consuming for broad scaled networks, the method of determining the balanced threshold level has been assessed for small scaled networks. However, this method can also be applied for broad scaled networks conceptually and the corresponding balanced threshold level can be calculated accordingly.

## ACKNOWLEDGEMENTS

The authors deeply acknowledge the support from DST, Govt. of India for this work in the form of FIST 2007 Project on "Broadband Wireless Communications" in the Department of ETCE, Jadavpur University.

## Authors

**Prasun Chowdhury** has completed his Masters in Electronics and Telecommunication Engineering from Jadavpur University, Kolkata, India in 2009. Presently he is working as Senior Research Fellow (SRF) in the Dept. of Electronics and Telecommunication Engineering, Jadavpur University. His current research interests are in the areas of Call Admission control and packet scheduling in IEEE 802.16 BWA Networks. He has authored international journals and conference papers. He is also a student member of IEEE and is presently the Secretary of the IEEE Students Branch of Jadavpur University.

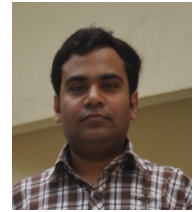

**Anindita Kundu** has completed her MTech in Distributed and Mobile Computing from Jadavpur University, Kolkata, India in 2010. Presently she is working as a senior research scholar in the Dept. of Electronics and Telecommunication engineering, Jadavpur University. Her current research interest is in the area of Broadband Wireless Communication and Cognitive Radio. She has also authored International Conferences and International Journals. She is also a student member of IEEE and is presently the Chairperson of IEEE Students Branch of Jadavpur University.

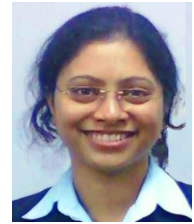

**Iti Saha Misra** received her PhD in Microstrip Antennas from Jadavpur University (1997). She is presently a Professor in the Department of Electronics and Telecommunication Engineering, Jadavpur University, India. Her current research interests include Mobility and Location Management, Next Generation Wireless Network Architecture and protocols, Call Admission control and packet scheduling in cellular and WiMAX networks, cognitive radio and green communication. She has authored more than 130 research papers in refereed Journal and International Conference and a book on Wireless Communications and Networks. She is an IEEE Senior Member and founder Chair of the Women In Engineering, Affinity Group, IEEE Calcutta Section.

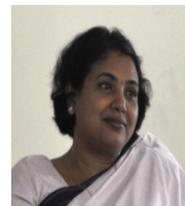

**Salil K. Sanyal** received his Ph.D from Jadavpur University, India (1990). He is currently a Professor in the Department of Electronics and Telecommunication Engineering, Jadavpur University. He has authored more than 140 Research Papers in refereed Journals and International/National Conference Proceedings and also co-authored the Chapter "Architecture of Embedded Tunable Circular Microstrip Antenna" in the book entitled "Large Scale Computations, Embedded Systems and Computer Security". He is a Senior Member of IEEE. His current research interests include Analog/Digital Signal Processing, VLSI Circuit Design, Wireless Communication and Tunable Microstrip Antenna.

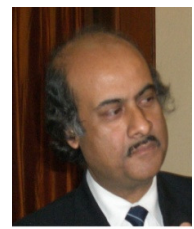